# ANNOTATING ANTISEMITIC ONLINE CONTENT. TOWARDS AN APPLICABLE DEFINITION OF ANTISEMITISM

Authors: Gunther Jikeli, Damir Cavar, Daniel Miehling

## ABSTRACT

Online antisemitism is hard to quantify. How can it be measured in rapidly growing and diversifying platforms? Are the numbers of antisemitic messages rising proportionally to other content or is it the case that the share of antisemitic content is increasing? How does such content travel and what are reactions to it? How widespread is online Jew-hatred beyond infamous websites and fora, and closed social media groups?

However, at the root of many methodological questions is the challenge of finding a consistent way to identify diverse manifestations of antisemitism in large datasets. What is more, a clear definition is essential for building an annotated corpus that can be used as a gold standard for machine learning programs to detect antisemitic online content. We argue that antisemitic content has distinct features that are not captured adequately in generic approaches of annotation, such as hate speech, abusive language, or toxic language.

We discuss our experiences with annotating samples from our dataset that draw on a ten percent random sample of public tweets from Twitter. We show that the widely used definition of antisemitism by the International Holocaust Remembrance Alliance can be applied successfully to online messages if inferences are spelled out in detail and if the focus is not on intent of the disseminator but on the message in its context. However, annotators have to be highly trained and knowledgeable about current events to understand each tweet's underlying message within its context. The tentative results of the annotation of two of our small but randomly chosen samples suggest that more than ten percent of conversations on Twitter about Jews and Israel are antisemitic or probably antisemitic. They also show that at least in conversations about Jews, an equally high number of tweets denounce antisemitism, although these conversations do not necessarily coincide.

# 1 Introduction

Online hate propaganda, including antisemitism, has been observed since the early days of popular internet usage.[1] Hateful material is easily accessible to a large audience, often without any restrictions. Social media has led to a significant proliferation of hateful content worldwide, by making it easier for individuals to spread their often highly offensive views. Reports on online antisemitism often highlight the rise of antisemitism on social media platforms (World Jewish Congress 2016; Community Security Trust and Antisemitism Policy Trust. 2019). Several methodological questions arise when quantitatively assessing the rise in online antisemitism: How is the rise of antisemitism measured in rapidly growing and diversifying platforms? Are the numbers of antisemitic messages rising proportionally to other content or is it also the case that the share of antisemitic content is increasing? Are antisemitic messages mostly disseminated on infamous websites and fora such as The Daily Stormer, 4Chan/pol or 8Chan/pol, Gab, and closed social media groups, or is this a wider phenomenon?

However, in addition to being offensive there have been worries that this content might radicalize other individuals and groups who might be ready to act on such hate propaganda. The deadliest antisemitic attack in U.S. history, the shooting at the Tree of Life Synagogue in Pittsburgh, Pennsylvania on October 27, 2018, and the shooting in Poway, California on April 27, 2019, are such cases. Both murderers were active in neo-Nazi social media groups with strong evidence that they had been radicalized there. White supremacist online radicalization has been a factor in other shootings, too, where other minorities, such as Muslims (Christchurch mosque shootings in March 2019) as well as people of Hispanic origin have been targeted (Texas shootings in August 2019). Extensive research and monitoring of social media posts might help to predict and prevent hate crimes in the future, much like what has been done for other crimes.[2]

Until a few years ago, social media companies have relied almost exclusively on users flagging hateful content for them before evaluating the content manually. The livestreaming video of the shootings at the mosque in Christchurch, which was subsequently shared within minutes by thousands of users, showed that this policy of flagging is insufficient to prevent such material from being spread. The use of algorithms and machine learning to detect such content for immediate deletion has been discussed but it is technically challenging and presents moral challenges due to censorial repercussions.

There has been rising interest in hate speech detection in academia, including antisemitism. This interest, including ours, is driven largely by the aim of monitoring and observing online antisemitism, rather than by efforts to censor and suppress such content. However, one of the major challenges is definitional clarity. Legal definitions are usually minimalistic and social media platforms' guidelines tend to be vague. "Abusive content" or "hate speech" is ill-defined, lumping together different types of abuse (Vidgen et al. 2019). We argue that antisemitic content has distinct features that are not captured adequately in more generic approaches, such as hate speech, or abusive or toxic language.

In this paper, we propose an annotation process that focuses on antisemitism exclusively and which uses a detailed and transparent definition of antisemitism. The lack of details of the annotation process and annotation guidelines that are provided in publications in the field have been identified as one of the major obstacles for the development of new and more efficient methods by the abusive content detection community (Vidgen et al. 2019, 85).

---

[1] The Simon Wiesenthal Center was one of the pioneers in observing antisemitic and neo-Nazis hate sites online, going back to 1995 when it found only one hate site (Cooper 2012).

[2] Yang et al. argue that the use of data from social media improves the crime hotspot prediction accuracy (D. Yang et al. 2018). Müller and Schwarz found a strong correlation between anti-refugee sentiment expressed on an AfD Facebook page and anti-refugee incidents in Germany (Müller and Schwarz 2017).

Our definition of antisemitism includes common stereotypes about Jews, such as "the Jews are rich," or "Jews run the media" that are not necessarily linked to offline action and that do not necessarily translate into online abuse behavior against individual Jews. We focus on (public) conversations on the popular social media platform Twitter where users of diverse political backgrounds are active. This enables us to draw from a wide spectrum of conversations on Jews and related issues. We hope that our discussion of the annotation process will help to build a comprehensive ground truth dataset that is relevant beyond conversations among extremists or calls for violence. Automated detection of antisemitic content, even under the threshold of "hate speech," will be useful for a better understanding of how such content is disseminated, radicalized, and opposed.

Additionally, our samples provide us with some indications of how users talk about Jews and related issues and how much of this is antisemitic. We used samples of tweets that are small but randomly chosen from all tweets in 2018 with certain keywords ("Jew*" and "Israel").

While reports on online antisemitism by NGOs, such as the Anti-Defamation League (Anti-Defamation League 2019; Center for Technology and Society at the Anti-Defamation League 2018), the Simon Wiesenthal Center (Simon Wiesenthal Center 2019), the Community Security Trust (Community Security Trust 2018; Stephens-Davidowitz 2019; Community Security Trust and Signify 2019), and the World Jewish Congress (World Jewish Congress 2018, 2016) provide valuable insights and resources, it has been noted that the fact that their data and methodology are concealed "places limits on the use of these findings for the scientific community" (Finkelstein et al. 2018, 2).

Previous academic studies on online antisemitism have used keywords and a combination of keywords to find antisemitic posts, mostly within notorious websites or social media sites (Gitari et al. 2015). Finkelstein et al. tracked the antisemitic "Happy Merchant" meme, the slur "kike" and posts with the word "Jew" on 4chan's Politically Incorrect board (/pol/) and Gab. They then calculated the percentage of posts with those keywords and put the number of messages with the slur "kike" in correlation to the word "Jew" in a timeline (Finkelstein et al. 2018). The Community Security Trust in association with Signify identified the most influential accounts in engaging with online conversations about Jeremy Corbyn, the Labour Party and antisemitism. They then looked at the most influential accounts in more detail (Community Security Trust and Signify 2019). Others, such as the comprehensive study on Reddit by the Center for Technology and Society at the Anti-Defamation League, rely on manual classification, but fail to share their classification scheme and do not use representative samples (Center for Technology and Society at the Anti-Defamation League 2018). One of the most comprehensive academic studies on online antisemitism was published by Monika Schwarz-Friesel (Schwarz-Friesel 2019). She and her team selected a variety of datasets to analyze how antisemitic discourse evolved around certain (trigger) themes and events in the German context through a manual in-depth analysis. This approach has a clear advantage over the use of keywords to identify antisemitic messages because the majority of antisemitic messages are likely more subtle than using slurs and clearly antisemitic phrases. The downside is that given the overwhelmingly large datasets of most social media platforms a preselection has to be done to reduce the number of posts that are then analyzed by hand.

Automatic classification using Machine Learning and Artificial Intelligence methods for the detection of antisemitic content is definitely possible, but, to our knowledge, relevant datasets and corpora specific to antisemitism have not been made accessible to the academic community, as of yet.[3] Various approaches focus on hate speech, racist or sexist content detection, abusive, offensive or toxic content, which include datasets and corpora in different languages, documented in the literature (e.g. (Anzovino, Fersini, and Rosso 2018; Davidson et al. 2017; Fortuna et al. 2019; Jigsaw 2017; Mubarak, Darwish, and Magdy 2017; Mulki et al. 2019; Nobata et al. 2016; Sanguinetti et al. 2018; Waseem and Hovy 2016), but the domain of antisemitism in Twitter seems to be understudied.

Keyword spotting of terms, such as the anti-Jewish slur "kike" or of images, such as the "Happy Merchant," might be sufficient on social media platforms that are used almost exclusively by White supremacists (Finkelstein et al. 2018). However, on other platforms they also capture messages that call out the usage of such words or images by other users. This is especially relevant in social media that is more mainstream, such as Twitter, as we show in detail below. Additionally, simple spotting of words such as "Kike" might lead to false results due to the possibility of varying meanings of these combinations of letters. Enrique García Martínez for example is a much-famed soccer player, known as Kike. News about him resulted in significant peaks of the number of tweets that contained the word "Kike" in our dataset from Twitter. However, even more sophisticated word level detection models are vulnerable to intentional deceit, such as inserting typos or change of word boundaries, which some users do to avoid automated detection of controversial content (Gröndahl et al. 2018; Warner and Hirschberg 2012).

Another way to identify antisemitic online messages is to use data from users and/or organizations that have flagged content as antisemitic (Warner and Hirschberg 2012). Previous studies have tried to evaluate how much of the content that was flagged by users as antisemitic has subsequently been removed by social media companies.[4] However, these methods then rely on classification by (presumably non-expert) users and it is difficult to establish how representative the flagged content is compared to the total content on any given platform.

This paper aims to contribute to reflections on how to build a meaningful gold standard corpus for antisemitic messages and to explore a method that can give us some indication of how widespread the scope of antisemitism really is on social media platforms like Twitter. We propose to use a definition of antisemitism that has been utilized by an increasing number of governmental agencies in the United States and Europe. To do so, we spell out how a concise definition could be used to identify instances of a varied phenomenon while applying it to a corpus of messages on social media. This enables a verifiable classification of online messages in their context.

---

[3] A widely used annotated dataset on racism and sexism is the corpus of 16k tweets made available on GitHub by Waseem and Hovy (Waseem and Hovy 2016), see http://github.com/zeerakw/hatespeech. Golbeck et al. produced another large hand coded corpus (of online harassment data) (Golbeck et al. 2017). Both datasets include antisemitic tweets, but they are not explicitly labeled as such. Warner and Hirschberg annotated a large corpus of flagged content from Yahoo! and websites that were pointed out to them by the American Jewish Congress as being offensive and they explicitly classified content as antisemitic (Warner and Hirschberg 2012). However, the corpus is not available to our knowledge.

[4] See "Code of Conduct on countering illegal hate speech online: One year after," published by the European Commission in June 2017, http://ec.europa.eu/newsroom/document.cfm?doc_id=45032, last accessed September 12, 2019).

## 2 DATASET

Our raw dataset is drawn from a ten percent sample of public tweets from Twitter via their Streaming API, collected by Indiana University's Network Science Institute's Observatory on Social Media (OSoMe). Twitter asserts that these tweets are randomly sampled. We cannot verify that independently, but we do not have any reason to believe otherwise. Tweets are collected live, on an ongoing basis, and then recorded (without images). We thus assume that the sample is indeed a representative sample of overall tweets. However, that does not mean that Twitter users or topics discussed within the sample are representative because users who send out numerous tweets will most likely be overrepresented, as well as topics that are discussed in many tweets.[5] For this paper, we use data from the year 2018.[6]

The overall raw sample includes 11,300,747,876 tweets in 2018. There were some gaps in the dataset from July 1 to 25, 2018, the number of tweets were only one percent instead of the usual ten percent.[7] OSoMe provides us with datasets of tweets with any given keyword in JSON format. For this paper, we used the queries "Israel" and "Jew*" for 2018. The query "Israel" includes all tweets that have the word Israel, followed or prefixed by space or signs, but not letters. The query "Jew*" includes all tweets that have the word Jew, followed by any letter or sign. The data includes the tweets' text, the sender's name, the date and time of the tweet, the number of retweets, the tweet and user ID, and other metadata. We had 3,427,731 tweets with the word "Jew*" from 1,460,075 distinct users in 2018 and 2,980,327 tweets with the word "Israel" from 1,101,373 distinct users.

This data was then fed into the Digital Method Initiative's Twitter Capture and Analysis Toolset (DMI-TCAT) for further analysis. Its open source code makes it transparent and allowed us to make changes which we used to tweak its function of producing smaller randomized samples from the dataset. We used this function to produce randomized samples of 400 tweets for both queries and then annotated these samples manually.

We divided the 2018 data into three portions because we did not want the month of July to be heavily underrepresented due to the gap in the raw data collection from July 1st to 25th.

---

[5] Indiana University Network Science Institute's Observatory on Social Media (OSoMe) collects a 10 percent sample of public tweets from Twitter via elevated access to their streaming API, since September, 2010. "An important caveat is that possible sampling biases are unknown, as the messages are sampled by Twitter. Assuming that tweets are randomly sampled, as asserted by Twitter, the collection does not automatically translate into a representative sample of the underlying population of Twitter users, or of the topics discussed. This is because the distribution of activity is highly skewed across users and topics and, as a result, active users and popular topics are better represented in the sample. Additional sampling biases may also evolve over time due to changes in the platform." (OSoMe, not dated, https://osome.iuni.iu.edu/faq/ last accessed July 23, 2019).

[6] We have also been collecting data from Twitter's Streaming API for a list of more than 50 keywords. Data analysts have speculated that it provides between 1 and 40 percent of all tweets with these keywords. However, a cross comparison of the two different dataset suggests that in some months, the Twitter API provides approximately 100 percent of the tweets with certain keywords. However, we have not used these datasets for this paper.

[7] The OSoMe project provides an interactive graph that shows the number of collected tweets per day, see https://osome.iuni.iu.edu/moe/tweetcount/stats/, last accessed September 12, 2019.

## OSoMe Highlights
Total Processed: 106,165,366,693

By Month

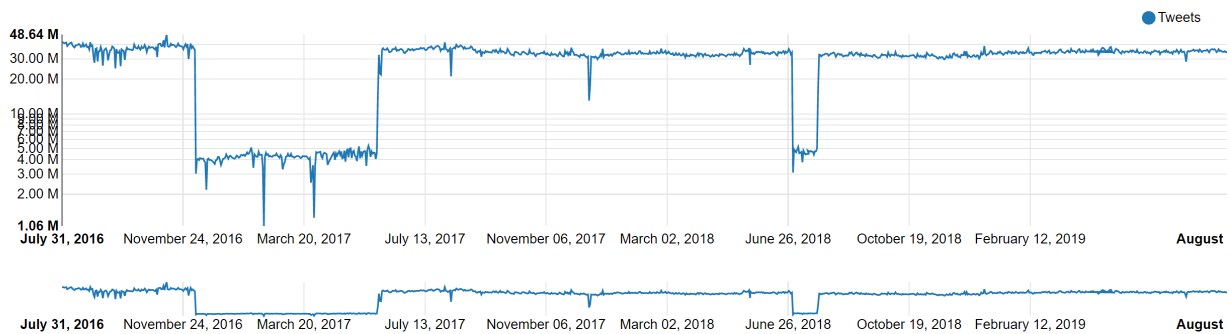

Graph 1: Number of Tweets Collected by OSoMe per Day. Source OSoMe

January 1st to June 30th, 2018 has a ten percent stream, July 1st to 25th, 2018, has a one percent stream, and July 26th to December 31, 2018, has ten percent as well. Thus, each sample has 198 tweets for the first period, 26 tweets from the second period and 176 tweets for the third period.[8]

The tweets' IDs allowed us to look at the tweets that were still live at the time of annotation on Twitter, which happened to be the majority of tweets. We were thus able to analyze the tweets within their context, including all images that were used, and also looking at previous tweets or reactions to the tweet.[9]

## 3 ANNOTATING TWEETS – DECIDING WHAT IS ANTISEMITIC AND WHAT IS NOT

Definitions of antisemitism vary and depend on their purpose of application (Marcus 2015). In legal frameworks the question of intent and motivation is usually an important one. Is a certain crime (partly) motivated by antisemitism? However, the intention of individuals is often difficult to discern in online messages and perhaps an unsuitable basis for definitions of online abuse (Vidgen et al. 2019, 82).

For our purposes, we are less concerned about motivation as compared to impact and interpretation of messages. Is a tweet likely to be interpreted in antisemitic ways? Does it transport and endorse antisemitic stereotypes and tropes?

---

[8] This method of sampling is the closest we could get to our goal to have a randomized sample with our keywords for the year 2018. A better method would have been to draw a randomized sample of 10 percent of our raw data from period one and three, to put it together with all data from period two (making it a one percent sample of all tweets for 2018) and to draw a randomized sample from that dataset. We have since found a method to overcome these challenges and we will use this method in future sampling.

[9] Vidgen et al. have pointed out that often long range dependencies exist and might be important factors in the detection of abusive content (Vidgen et al. 2019, 84). Yang et al. showed that augmenting text with image embedding information improves automatically identifying hate speech (F. Yang et al. 2019).

## INTENT OR IMPACT?

The motivation and intent of the sender is not relevant for us because we do not want to determine if the sender is antisemitic but rather if the message's content is antisemitic. A message can be antisemitic without being intentional. This can be illustrated with a tweet by a Twitter bot that sends out random short excerpts from plays by William Shakespeare. We came across a tweet that reads "*Hebrew, a Jew, and not worth the name of a Christian.*" sent out by the user "*IAM_SHAKESPEARE,*" see image 1 below. A look at the account confirms that this bot randomly posts lines from works of Shakespeare every ten minutes. Antisemitic intent of the bot and its programmer can almost certainly be ruled out. However, some of the quotes might carry antisemitic tropes, possibly the one mentioned above because it suggests a hierarchy between Jews and Christians and inherit negative character traits among Jews.

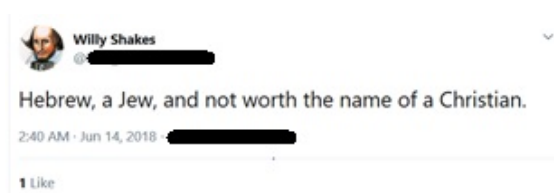

Image 1: "Shakespeare"

We thus look at the message itself and what viewers are likely to take away from it. However, it is still necessary to look at the overall context to understand the meaning and likely interpretations of it. Some, for example, might respond to a tweet with exaggerated stereotypes of Jews as a form of irony to call out antisemitism. However, the use of humor, irony and sarcasm does not necessarily mean that certain stereotypes are not disseminated (Vidgen et al. 2019, 83).

Lastly, when examining the impact of a tweet, we only assessed the potential for transmitting antisemitism. Some tweets in our dataset expressed negativity towards other groups while also expressing animus towards Jews. Our annotation approach for this study focused on the components that pertained to antisemitism only, even while the authors of this paper acknowledge that animus for other groups is often related to antisemitism.

## FINDING THE RIGHT DEFINITION

We use the most widely used definition of contemporary antisemitism, the Working Definition of Antisemitism by the International Holocaust Remembrance Alliance (IHRA). [10] This non-legally binding definition was created in close cooperation with major Jewish organizations to help law enforcement officers and intragovernmental agencies to understand and recognize contemporary forms of antisemitism. Its international approach, its focus on contemporary forms, and the many examples that are included in the definition make it particularly useful for annotating tweets. However, many parts of the definition need to be spelled out to be able to use it as a standardized guideline for annotating tweets. For example, the definition mentions "classic stereotypes" and "stereotypical allegations about Jews as such," without spelling out what they are. Spelling out the key stereotypes is necessary due to the vast quantity of stereotypes that have arisen historically, many of which are now hard to recognize outside of their original context. We did a close reading of the definition allowing for inferences that can clarify certain grey zones in the annotation. We also consulted the major literature on "classic" antisemitic stereotypes and stereotypical allegations to list the stereotypes and accusations against Jews that are considered to be "classical" antisemitism. The full text of the definition and our inferences can be found in Annex I and Annex II. Both texts served as the basis for our annotation.

---

[10] The IHRA Working Definition has been adopted by more than a dozen governments to date. For an updated list see https://www.holocaustremembrance.com/working-definitions-and-charters. It has also been endorsed as a non-legal educational tool by the United Nation's Special Rapporteur on freedom of religion or belief, Ahmed Shaheed (Office of the United Nations High Commissioner for Human Rights 2019).

## WHO ARE THE ANNOTATORS?

Scholars have argued that the various forms of antisemitism and its language can transform rapidly and that antisemitism is often expressed in indirect forms (Schwarz-Friesel and Reinharz 2017). Members of targeted communities, that is Jews in the case of antisemitism, are often more sensitive in detecting the changing language of hatred against their community. While monitoring bigotry, the perspective of targeted communities should be incorporated. Other studies on hate speech, such as the Anti-Defamation League's study on hate speech on Reddit, used an "intentionally-diverse team to annotate comments as hate or not hate" (Center for Technology and Society at the Anti-Defamation League 2018, 17). The annotators in our project were graduate students who had taken classes on antisemitism and undergraduate students of Gunther Jikeli's course "Researching Antisemitism in Social Media" at Indiana University in Spring 2019. All annotators had participated in extensive discussions of antisemitism and were familiarized with the definition of antisemitism that we use, including our inferences. Although our team of annotators had Jewish and non-Jewish members, we aimed for a strict application of our definition of antisemitism that incorporates perspectives from major Jewish organizations. Within a relatively small team, individual discrepancies were hypothesized to be dependent upon training and attentiveness in the application of the definition rather than on an annotator's background. One of the four annotators of the two samples that are discussed in this paper is Jewish. They classified slightly less tweets as antisemitic than her non-Jewish counterpart, see annotation results in table 1 below.

## GREY ZONES IN THE ANNOTATION

Although we concentrate on the message itself and not on the motivation of the sender, we still need to examine the context in which the tweet is read, that is, the preceding tweets if situated in a thread. Reactions to it can also provide information on how a particular tweet has been interpreted. We are looking at all the information that the reader is likely to see, and which will be part of the message the reader gets from a given tweet. We also examine embedded images and links.[11]

We consider that posting links to antisemitic content without comment a form of disseminating content and therefore, an antisemitic message. However, if the user distances themselves from such links, in direct or indirect ways, using sarcasm or irony which makes it clear that there is disagreement with the view or stereotype in question, then the message is not considered antisemitic. Part of the context that the reader sees is the name and profile image of the sender. Both are visible while looking at a tweet. A symbol, such as a Nazi flag or a dangerous weapon might sway the reader to interpret the message in certain ways or might be antisemitic in and of itself as is the case with a Nazi flag.

When it comes to grey zones, we erred on the side of caution. We classified tweets as antisemitic if they add antisemitic content, or if they directly include an antisemitic message, such as retweeting antisemitic content, or quoting antisemitic content in an approving manner.

Endorsement of antisemitic movements, organizations, or individuals are treated as symbols. If they stand for taking harmful action against Jews, such as the German Nazi party, the Hungarian Arrow Cross, Hamas, Hitler, well-known Holocaust deniers, Father Coughlin, Mahmoud Ahmadinejad, David Duke, or The Daily Stormer, then direct endorsement of such individuals, organizations, or movements are considered equivalent to calls for harming Jews and therefore antisemitic. If they are known for their antisemitic action or words but also for other things, then it depends on the context and on the degree to which they are known to be antisemitic. Are they endorsed in a way that the antisemitic position for which these organizations or individuals stand for is part of the message? The British Union of Fascists, Mussolini, Hezbollah, or French comedian Dieudonné are likely to be used in such a way, but not necessarily so.

---

[11] Annotators were instructed to spend not more than five minutes on links, significantly more than the average smartphone user spends on longer news items (PEW Research Center 2016).

Despite our best efforts to spell out as clearly as possible what constitutes antisemitic messages, there will remain room for interpretation and grey zones. The clarifications in Annex II are the results of our discussions about tweets that we classified differently in a preliminary study.

### ANNOTATION SCHEME

How did we annotate the samples? We ran a script to separate deleted tweets from tweets that are still live and focused on those. However, some tweets were deleted between the time that we ran the script and annotated our samples. The first annotation point is therefore an option to mark if tweets are deleted. The second point of annotation provides an option to indicate if the tweet is in a foreign language.

Our main scheme, to decide if tweets are antisemitic or not, explicitly according to the IHRA Definition and its inferences, was coded on a five-point scale from -2 to 2 where:

-2 = Tweet is antisemitic (confident)
-1 = Tweet is antisemitic (not confident)
0 = Tweet is not comprehensible
1 = Tweet is not antisemitic (not confident)
2 = Tweet is not antisemitic (confident).

The next annotation point gives the annotators the option to disagree with the IHRA Definition with respect to the tweet at hand. This gives the annotators the opportunity to classify something as antisemitic which does not fall under the IHRA Definition or vice versa and might help to reduce personal bias if the annotators disagree with the IHRA Definition.

We also asked annotators to classify tweets with respect to the sentiments that tweets evoke towards Jews, Judaism, or Israel independently from the classification of antisemitism.[12] This was done on a five-point scale as well from 1-5 where:

-2 = Tweet is very negative towards Jews, Judaism, or Israel
-1 = Tweet is negative towards Jews, Judaism, or Israel
0 = Tweet has neutral sentiment or is not comprehensible
1 = Tweet is positive towards Jews, Judaism, or Israel
2 = Tweet is very positive towards Jews, Judaism, or Israel.

A categorization such as this helps annotators to stick to the IHRA Definition. If a tweet is negative towards Jews, Judaism, or Israel but it does not fit the definition of the IHRA Definition, then annotators still have a way to express that. While most antisemitic tweets will be negative towards Jews or Israel there are oftentimes positive stereotypes such as descriptions of Jews as inherently intelligent or good merchants. On the other hand, some descriptions of individual Jews, Judaism, or Israel can be negative without being antisemitic. The sentiment scale might also be useful for further analysis of the tweets. Lastly, annotators can leave comments on each tweet.

It took the annotators two minutes on average to evaluate each tweet.[13]

Graphs 2 and 3 show the timelines for the number of tweets per day that include the word Jew* and Israel.

---

[12] Sentiment have been identified as an important indicator for hate speech or "toxic content" (Brassard-Gourdeau and Khoury 2019).
[13] We wrote an annotation program that facilitates the annotation process by pulling up the tweets and making choices clickable for annotators. Unfortunately, the program was not fully operational and so we reverted to spreadsheets that link to Twitter.

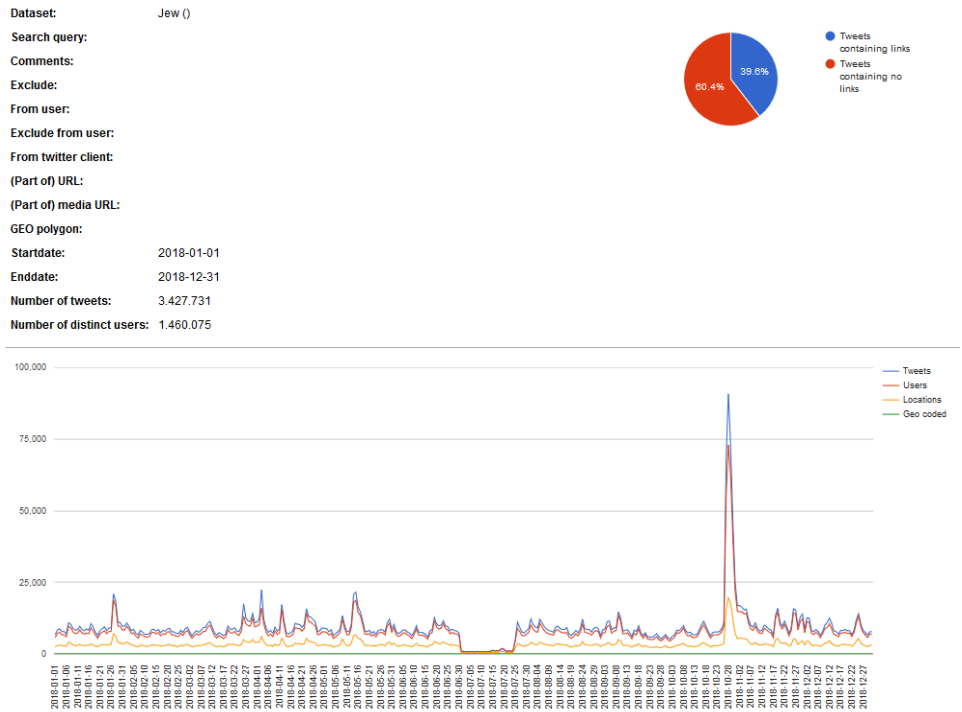

Graph 2: Number of Tweets in 2018 per Day that have the Word Jew* (10 percent randomized sample), graph by DMI TCAT

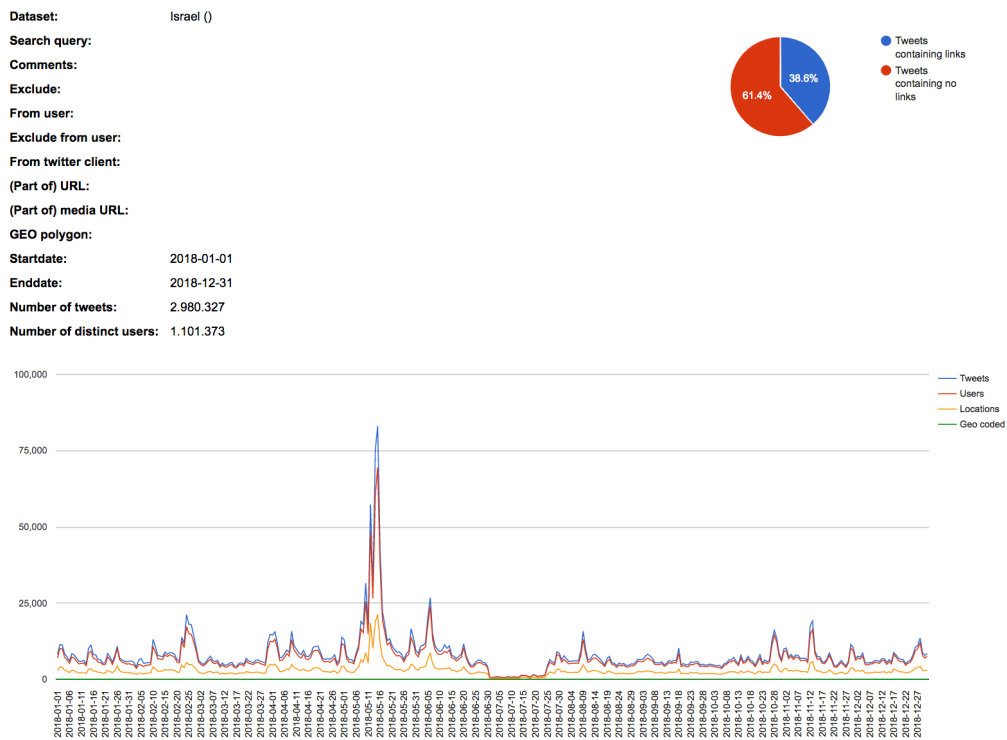

Graph 3: Number of Tweets 2018 per Day that have the Word Israel (10 percent randomized sample), graph by DMI TCAT

# 4 PRELIMINARY RESULTS

## PEAKS OF CONVERSATIONS ABOUT JEWS AND ISRAEL

We draw from a dataset that collects 10 percent of all tweets, randomly sampled, see description of the dataset above. This includes 3,427,731 tweets from 1,460,075 different users that have the three letters JEW in this sequence, including Jewish, Jews, etc., in 2018. Our dataset also contains 2,980,327 tweets from 1,101,371 different users that have the word Israel in it (not including derivates, such as Israelis). From July 1, 2018, to the first part of July 25, 2018, the overall dataset included only one percent of all tweets. This resulted in a drop in the number of tweets containing our keywords during that period.

The highest peaks of tweets with the word Jew* can be linked to offline events. By far, the highest peak is shortly after the shootings at the Tree of Life synagogue in Pittsburgh, October 27, 2018. The second highest peak is during Passover, one of the most important Jewish holidays. British Labour opposition leader Jeremy Corbyn's visit to a Seder event, April 3, at the controversial Jewish group "Jewdas" was vividly discussed on social media, including charges of antisemitism. The third peak can be found at the time when the U.S. Embassy was moved to Jerusalem. The fourth highest peak is on Holocaust Memorial Day, January 27. The fifth highest relates to a protest outside British Parliament against antisemitism within the Labour Party, March 26.

The five highest peaks of tweets containing the term Israel can also be linked to offline events. The highest peak on May 15, 2018, is the date when the U.S. Embassy was moved to Jerusalem, followed by the second highest peak, May 12, on the day when Netta Barzilai won the Eurovision Song Contest 2018 for Israel with her song "Toy." The peak on May 10 relates to the Iranian rocket attack against Israel from Syrian territory and the response by Israeli warplanes. The peak on June 6 seems to be related to the successful campaign to cancel a friendly soccer match between Argentina and Israel and to the eruption of violence at the Israeli-Gazan border. The fifth highest peak in 2018 with the word "Israel" does not relate to the country Israel but to the last name of a law enforcement officer in Broward County, Florida. Sheriff Scott Israel came under scrutiny for his role at the Parkland High School shooting and important details were revealed on February 23.

## PERCENTAGES OF ANTISEMITIC TWEETS IN SAMPLES

We drew randomized samples of 400 tweets for manual annotation from 2017 and 2018. In this paper, we discuss the annotation of two samples from 2018 by two annotators, each. One sample is a randomized sample of tweets with the word "Jew*" and the other with the word "Israel." Previous samples with these terms included approximately ten percent that were antisemitic. Assuming that the proportion of antisemitic tweets is no larger than twenty percent, we can calculate the margin of error as up to four percent for the randomized sample of 400 tweets for a 95 percent confidence level. [14] However, we discarded all deleted tweets and all tweets in foreign languages, which led to a significantly reduced sample size. In the sample of tweets with the word "Jew*" we also discarded all tweets that contained the word "jewelry" but not "Jew". [15] This resulted in a sample of 172

---

[14] n = p(1-p)(Z/ME)^2 with p= True proportion of all tweets containing antisemitic tweets; Z= z-score for the level of confidence (95% confidence); and ME= margin of error.

[15] We also discarded tweets with a misspelling of "jewelry," that is "jewerly" and "jewery." 128 tweets (32 percent) were thus discarded. The large number of tweets related to jewelry seem to be spread out more evenly throughout the year. They

tweets with the word "Jew*" and a sample of 247 tweets with the word "Israel." [16] The calculated margin of error for a sample of 172 tweets is six percent and for a sample of 247 tweets it is five percent. However, there might be bias in discarding deleted tweets because the percentage of antisemitic tweets might be higher among deleted tweets. Although the text and metadata of deleted tweets was captured, we could not see the tweets in their context, that is, with images and previous conversations. Looking through the texts of deleted tweets shows that many but far from all deleted tweets are likely antisemitic.

Some annotation results can be seen in table 1 below.

|  | "Jew*" 2018 Sample Annotator B | | "Jew*" 2018 Sample Annotator G | | "Israel" 2018 Sample Annotator D | | "Israel" 2018 Sample Annotator J | |
|---|---|---|---|---|---|---|---|---|
| **Sample size without deleted tweets and tweets in foreign language (and without "Jewelry" tweets)** | 172 | | 172 | | 247 | | 247 | |
| **Confident antisemitic** | 10 | 5.8% | 9 | 5.2% | 16 | 8.2% | 11 | 5.6% |
| **Probably antisemitic** | 21 | 12.2% | 12 | 7.% | 15 | 6.1% | 12 | 4.9% |
| *SUM (probably) antisemitic* | *31* | *18.%* | *21* | *12.2%* | *31* | *14.3%* | *23* | *10.5%* |
| | | | | | | | | |
| **Calling out antisemitism** | 25 | 14.5% | 36 | 18.5% | 12 | 6.2% | 4 | 2.1% |

Table 1: Annotation Results of Two Samples

The first annotator of the sample with the word "Jew*" classified eighteen percent of the tweets as antisemitic or probably antisemitic, while the second annotator classified twelve percent as such. The first annotator of the sample with the word "Israel" classified fourteen percent of the tweets as antisemitic or probably antisemitic, while the second annotator classified eleven percent as antisemitic/probably antisemitic. Interestingly, a high number of tweets (fifteen and nineteen percent) with the word "Jew*" were found to be calling out antisemitism. Tweets calling out antisemitism were significantly lower within the sample of "Israel" tweets (six and two percent).

The discrepancies in the annotation were often a result of a lack of understanding of the context, in addition to lapses in concentration and different interpretations of the respective tweets. Different interpretations of the definition of antisemitism seem to have been relatively rare. We discussed different interpretations of the definition in trial studies, which led to clarification that is now reflected in our inferences of the IHRA Working Definition, see Annex II.

---

did not result in noticeable peaks. However, future queries should exclude tweets related to jewelry to avoid false results in peaks or other results related to metadata.

[16] The annotators did not do the annotation at the same time. Some tweets were deleted during that time. For better comparison we only present the results of the annotation of tweets that were live during both annotations.

## EXAMPLES OF TWEETS THAT ARE DIFFICULT TO FULLY UNDERSTAND

Further investigation and discussion between annotators can help to understand the meaning of messages that are difficult to understand. This can lead to a re-classification of tweets. The three examples below appear less likely to be read in antisemitic ways than previously thought. Other cases went the opposite way.

A good example for the difficulty of understanding the context and its message is the tweet "*@realMatMolina @GothamGirlBlue There is a large Jewish population in the neighborhood and school*" by user "*WebMaven360*" with the label "*Stephen Miller – Secretary of White Nationalism*". The text itself without any context is certainly not antisemitic. However, annotator B classified it as "probably not antisemitic," perhaps because of the user's label. The tweet responds to another tweet that reads: "*Few facts you may not know about the Parkland shooter: • He wore a Make America Great Again hat • Had a swastika carved into his gun • Openly expressed his hate for Muslims, Jews, and black people. Why weren't these things ever really talked about?*" The response "*There is a large Jewish population in the neighborhood and school*" can be read as a Jewish conspiracy that suppresses this kind of information in the media. This would be covered by the definition's paragraph 3.1.2, see Annex I. Annotator G classified the tweet as antisemitic. However, another, and probably more likely reading of the tweet sequence is that both users are critical of Trump and that the second tweet is simply providing additional evidence in support of the first tweet. Specifically, the second tweet is suggesting that the Parkland shooting was racially motivated, that it was not a random case of school violence, but rather aimed at a largely Jewish student body. Looking at other tweets of "*Stephen Miller – Secretary of White Nationalism*" and their followers reveals that they are in fact very critical of Stephen Miller and the Trump administration and often denounce racial bigotry. It is therefore likely that the message of this tweet did not transmit any antisemitic connotations to the readers.

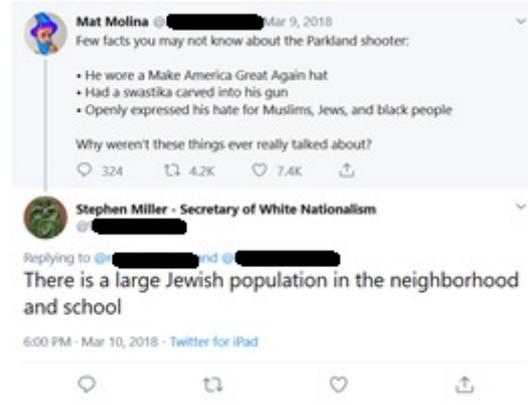

Image 2: "There is a large Jewish population…"

A careful reading is necessary and closer investigation can often clarify if there is an antisemitic message or not. In other cases it remains unclear. A case in point is a comment about Harvey Weinstein and his interview with Taki Theodoracopulos in the Spectator, July 13, 2018.

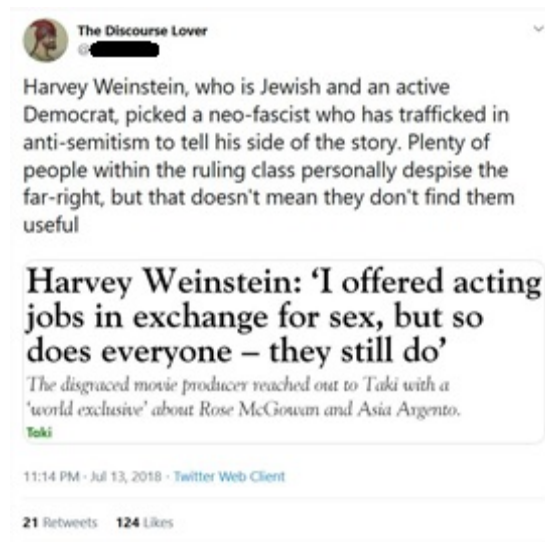

Image 3: "Harvey Weinstein, who is Jewish…"

The tweet comments on the headline of the interview with Weinstein and on the fact that the interviewer is a well-known figure of the extreme right.[17] The tweet reads "*Harvey Weinstein, who is Jewish and an active Democrat, picked a neo-fascist who has trafficked in anti-semitism to tell his side of the story. Plenty of people within the ruling class personally despise the far-right, but that doesn't mean they don't find them useful.*" The first sentence which identifies Weinstein as Jewish and a democrat is not antisemitic. The second sentence however identifies Weinstein also as a member of "the ruling class" who "uses" the far-right and even a "neo-fascist" for their own purposes. It is unclear what makes him a member of "the ruling class" and if his Jewishness is seen as a factor for this. The two sentences together, however, can be interpreted as the old antisemitic stereotype of Jews being or influencing "the ruling class" and using all means to push their political agenda. This is covered by the definition's paragraph 3.1.2. Both annotators classified the tweet as "probably antisemitic." However, there is also a non-antisemitic reading of the tweet sequence whereby Weinstein is part of the ruling class by virtue of his wealth and influence in an important industry. Alluding to Weinstein's Jewishness would thus serve to further illustrate elites' supposed willingness to use even their worst enemies if it helps them stay in power. The reference to Weinstein's Jewish heritage can thus be read as secondary in importance, used only to highlight the extreme lengths an elite would go to stay in power. At the same time, any reader with animus to Jews may elevate the importance of Weinstein's Jewish heritage regardless of the author's intent.

Some tweets that contain antisemitic stereotypes can be read as such or as calling out antisemitism by exaggerating it and mocking the stereotypes. User "*Azfarovski*" wrote "*If you listen carefully, Pikachu says "Pika pika Pikachu" which sounds similar to "Pick a Jew" which is English for "Yahudi pilihan saya" Allahuakhbar. Another one of the ways they are trying to corrupt the minds of young Muslims*." "*Azfarovski*" seems to be the genuine account of Azfar Firdaus, a Malaysian fashion model.[18] The account has more than 45,000 followers. The particular tweet was retweeted more than 4,000 times and was liked by almost the same number of users. The tweet contains a particular absurd antisemitic conspiracy theory, a conspiracy theory however, that closely resembles widespread rumors about Pokemon in the Middle East that even led to a banning of Pokemon in Saudi Arabia and the accusation of promoting "global Zionism" and Freemasonry.[19] These kind of rumors are covered by paragraphs 3.0 and 3.1.2 of the Working Definition of Anti-Semitism and the accusation that Jews allegedly conspire to wage war against "the Muslims" is a classic antisemitic stereotype within Islamist circles, see paragraph on "Jewish crimes" in Annex II. Both annotators classified the tweet as antisemitic.

However, there is also the possibility that this tweet ridicules antisemitic conspiracy theories and thereby calls out antisemitism. How is the tweet read, what is the context? What is the evidence for it being anti-antisemitic instead of it transmitting an antisemitic conspiracy theory?

---

[17] Taki Theodoracopulos founded The American Conservative magazine with Pat Buchanan and Scott McConnell, wrote in support of the Greek ultranationalist political party Golden Dawn, made Richard Spencer an editor of his online magazine and he was accused of antisemitism even in the Spectator, March 3, 2001.

[18] Azfarovski used a name in Arabic (see screenshot) that reads "albino broccoli." He has since changed that twitter handle several times.

[19] Pokemon was banned in Saudi Arabia with a fatwa in 2001. The fatwa accused Pokémon of promoting the Shinto religion of Japan, Christianity, Freemasonry and "global Zionism." The Times of Israel, July 20, 2016, "Saudi revives fatwa on 'Zionism-promoting' Pokemon," https://www.timesofisrael.com/saudi-fatwa-on-zionist-pokemon-for-promoting-evolution/

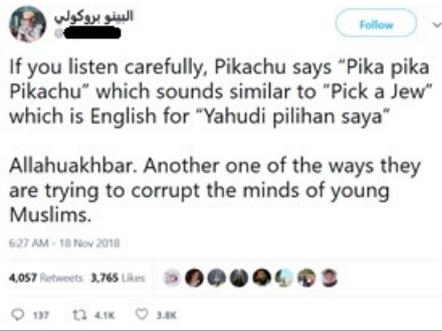

Image 4: "If you listen carefully, Pikachu says…"

"*Azfarovski's*" other tweets and discussions in his threads are rarely about Jews or anything related. There are some allusions to conspiracy theories with Illuminati, but they are rare and made (half-) jokingly. We did not find any tweet in which he distanced himself from conspiracy theories or bigotry against Jews. However, back in 2016, "*Azfarovski*" wrote a similar tweet. He commented on a discussion in which another user questioned the alleged connection between Pokemon and Jews: "*Pikachu is actually pronounced as Pick-A-Jew to be your friend. Omg. OMG ILLUMINATI CONFIRMED.*" This is extremely suggestive and might have been read as satire. While the user's intention remains unclear, how do his followers interpret the tweet?

The direct responses to the tweet show a mixed picture.[20] The response "*lowkey upset that i was born too late to truly appreciate the massive waves of Pokemon conspiracies back in the late 90s and early 00s,*" shows that its author, "*TehlohSuwi*" dismisses this as a, perhaps funny, conspiracy theory and does not take it seriously. Others however responded with memes of disbelief such as the images below or various forms of disagreement, such as user "*namuh*" who said "*What kind of shit is this? Equating Pika pika Pikachu to "pick a Jew" is outta this world! [...]*". This suggests that they took the message at face value but disagreed.

---

[20] Many of the responses were in Indonesian. The examples presented here were in English.

User "*medicalsherry*" was unsure: "*This is sarcasm,right?*" Other users did not object to the conspiracy theory but just to the alleged impact of it. "*Not for who don't really mind about it.. unless their mind is really easy and wanted to be corrupted..*" replied user "*ash_catrina*".

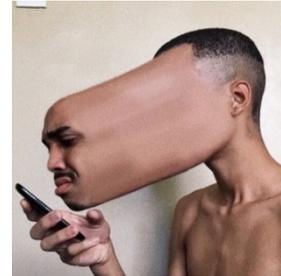

Image 5: User "shxh's" response to "Azfarovski's" tweet, 18 Nov 2018

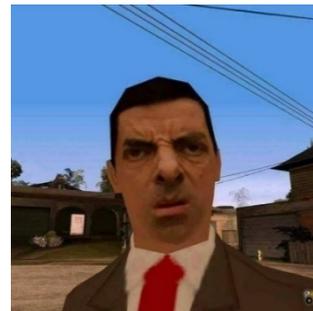

Image 6: User "monoluque's" response to "Azfarovski's" tweet, 18 Nov 2018

Going through the profiles and tweet histories of the 4000+ users who retweeted the tweet in question also provides information about how the tweet was perceived. The large majority of them has neither a history of antisemitic conspiracy theories nor of calling them out. However, some of them show some affinity to conspiracy theories, e.g. about Illuminati.

Thus, the tweet evoked antisemitic connotations at least in some of the readers even if it cannot be established whether the disseminator endorses this fantasy or they are merely mocking it.

## DIFFERENCES BETWEEN ANNOTATORS

Disagreement between annotators are due to a number of reasons, including lack of context and different opinions of the definition of the bias in question (Waseem and Hovy 2016, 89–90). In our tweets including the word "Jew*" annotator B classified more tweets as "probably antisemitic" than annotator G. This is partly because annotator B is not sufficiently familiar with ways to denounce antisemitism and with organizations who do so. For example, annotator B classified a tweet by MEMRI, the Middle East Media Research Institute, an organization that tracks antisemitism in media from the Middle East, as antisemitic. This tweet documents a talk that might be seen as antisemitic, see screenshot image 7 below. However, the tweet calls this out and annotator G correctly classified this as calling out antisemitism and confidently not antisemitic.

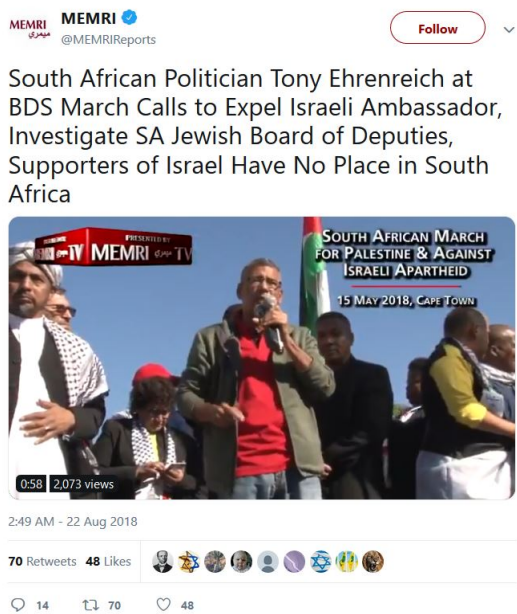

Image 7: "South African Politician…"

Different classifications of other tweets might be due to unclear messages. User "*Scrooched*" tweeted "*Seems like Jared is a Jew/Nazi Hitler would've been proud of. We know that Miller is a white supremacists & that #IdiotInChief is listening to those dangerous to our democracy fools. Wouldnt it be nice if we had @POTUS who had a brain and believed our rule of law #TrumpCrimeFamily.*" Saying that Jared Kushner is a Nazi or "a Jew Hitler would've been proud of" can be seen as a form of demonizing Kushner and even of downplaying the Nazi ideology in which it was not conceivable that Hitler would have been "proud" of any Jew. Annotator B therefore classified the tweet as probably antisemitic while annotator G did not see enough evidence for a demonization of a Jewish person or for denying the "intentionality of the genocide of the Jewish people at the hands of National Socialist Germany" (Working Definition, paragraph 3.1.4) and classified the tweet as "probably not antisemitic."

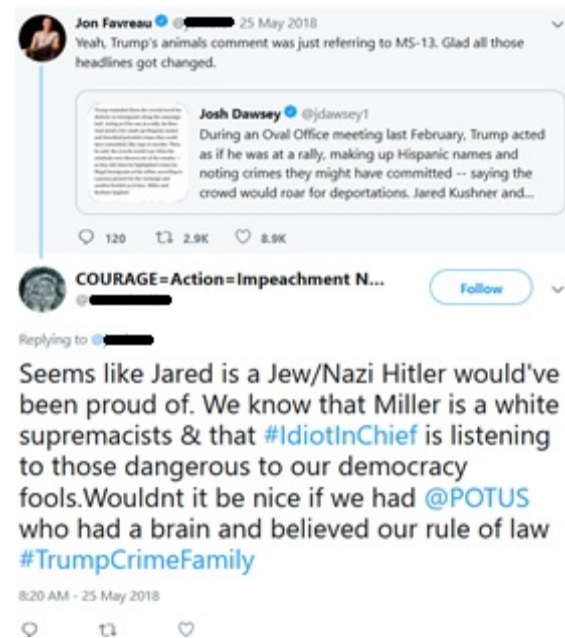

Image 8: "Seems like Jared is a Jew/Nazi…"

The discrepancies between the two annotators for the "Israel" sample were bigger than between the annotators for the "Jew*" sample but the reasons were similar. Annotator J classified a tweet by "syria-updates" (see screenshot image 9) as probably not antisemitic while annotator D classified it as antisemitic. The tweet contains a link to an article on the "Syria News" website entitled "Comeuppance Time as Syrian Arab Army Strikes Israel in the Occupied Golan Heights," from May 12, 2018. The caption under the embedded image in the tweet suggests that this might be antisemitic by using the phrase "*MI6/CIA/Israel loyalist media made claims that [...]*."However, it is only when reading the linked article itself it becomes clear that antisemitic tropes are disseminated. The article states that "*the media*" are reading "*directly from the Israeli script*." The classic antisemitic image of Jews controlling the media is thus used to characterize Israel, which is covered by paragraph 3.1.9 of the IHRA Working Definition. The discrepancy in the annotation seems to stem from one annotator spending time reading the linked article while the other did not.

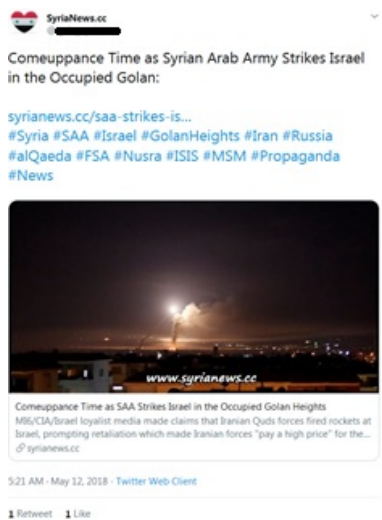
Image 9: "SyrianNews.cc"

A tweet that both annotators agreed was spreading an antisemitic message was a retweet that was originally sent out by the user "*BDSmovement*." It included a link to a short video in which former United Nations Special Rapporteur Richard Falk proclaimed that Israel was an Apartheid state and called for the Jewish State to cease to exist, which is covered by paragraph 3.1.7 of the IHRA Definition, that is "Denying the Jewish people their right to self-determination, e.g., by claiming that the existence of a State of Israel is a racist endeavor."

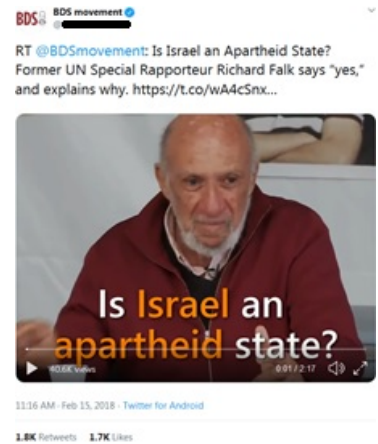
Image 10: "BDS movement"

## 5 Discussion

We applied the IHRA Working Definition of Antisemitism to tweets by spelling out some of the inferences necessary for such an application. In particular we expand on the classic antisemitic stereotypes that are not all named in the Working Definition by going through scholarly works that have focused on such classic antisemitic stereotypes. The IHRA Definition and the documented inferences are sufficiently comprehensive for the annotation of online messages on the mainstream social media platform Twitter. However, additional antisemitic stereotypes and antisemitic symbols (images, words, or numbers) exist in circles that we have not yet explored, and some will be newly created. These can be added to the list as needed.

In evaluating messages on social media that connect people around the world, partly anonymously, it does not make sense to evaluate intent. The messages, which often get retweeted by other users, should be evaluated for their impact in a given context. The question is if the message disseminates antisemitic tropes and not if that was the intention.

The annotation of tweets and comparison between annotators showed that annotators need to be a) highly trained to understand the definition of antisemitism, b) knowledgeable about a wide range of topics that are discussed on Twitter, and, perhaps most importantly, c) be diligent and detail-oriented, including with regard to embedded links and the context. This is particularly important to distinguish antisemitic tweets from tweets that call out antisemitism which they often do by using irony. This confirms findings of another study that relates low agreement between annotators of hate speech to the fact that they were non-experts (Fortuna et al. 2019, 97). Discussion between qualified annotators in which they explain the rationale for their classification is likely to result in better classification than using statistical measures across a larger number of (less qualified) annotators.

An analysis of the 2018 timeline of all tweets with the word "Jew*" (3,427,731 tweets) and "Israel" (2,980,327 tweets), drawn from the ten percent random sample of all tweets, show that the major peaks are correlated to discussions of offline events. In our representative samples of live tweets on conversations about Jews and Israel we found relatively large numbers of tweets that are antisemitic or probably antisemitic, between eleven and fourteen percent in conversations including the term "Israel" and between twelve and eighteen percent in conversations including the term "Jew,*" depending on the annotator. It is likely that there is a higher percentage of antisemitic tweets within deleted tweets. However, in conversations about Jews the percentage of tweets calling out antisemitism was even higher (between fifteen and nineteen percent depending on the annotator). This was not the case in conversations about Israel where only a small percentage of tweets called out antisemitism (two to six percent depending on the annotator). Antisemitism related to Israel seems to be highlighted and opposed less often than forms of antisemitism that are related to Jews in general. These preliminary findings should be examined in further research.

Our study does not provide an annotated corpus that can serve as a gold standard for antisemitic online message, but we hope that these reflections might be helpful towards this broader goal.

# 6 ACKNOWLEDGMENTS

This project was supported by Indiana University's New Frontiers in the Arts & Humanities Program. We are grateful that we were able to use Indiana University's Observatory on Social Media (OsoMe) tool and data (Davis et al. 2016). We are also indebted to Twitter for providing data through their API. We thank David Axelrod for his technical role and conceptual input throughout the project, significantly contributing to its inception and progress. We thank the students of the course "Researching Antisemitism in Social Media" at Indiana University in Spring 2019 for the annotation of hundreds of tweets and discussions on the definition and its inferences, namely Jenna Comins-Addis, Naomi Farahan, Enerel Ganbold, Chea Yeun Kim, Jeremy Levi, Benjamin Wilkin, Eric Yarmolinsky, Olga Zavarotnaya, Evan Zisook, and Jenna Solomon. The funders had no role in study design, data collection and analysis, decision to publish, or preparation of the manuscript.

# ANNEX I: IHRA DEFINITION OF ANTISEMITISM

The International Holocaust Remembrance Alliance (IHRA), an organization with 31 member countries, including the United States, Canada, and most EU countries, adopted a non-legally binding working definition of antisemitism in 2016, which is a slightly modified version of the Working Definition that was previously used unofficially by the European Union Monitoring Centre on Racism and Xenophobia (EUMC) and its successor organization the EU Agency for Fundamental Rights (FRA). The definition has since been adopted by nine country governments and numerous governmental and non-governmental bodies.[21] In December 2018, the European Council called on all EU countries to also adopt the definition. It is thus by now the most widely accepted definition of antisemitism. The definition was "created in response to a perceived need by police officers and the intergovernmental agencies to understand the forms and directions that antisemitism now takes, and after consideration of many drafts by a wide range of Jewish and non-Jewish specialists." (Whine 2018, 16).

For the sake of clarity, we gave the definition section labels. Otherwise, the text below in square brackets is the unaltered text of the IHRA Working Definition.

*[1.0 "Antisemitism is a certain perception of Jews, which may be expressed as hatred toward Jews. Rhetorical and physical manifestations of antisemitism are directed toward Jewish or non-Jewish individuals and/or their property, toward Jewish community institutions and religious facilities."*

*2.0 To guide IHRA in its work, the following examples may serve as illustrations:*

*3.0 Manifestations might include the targeting of the state of Israel, conceived as a Jewish collectivity. However, criticism of Israel similar to that leveled against any other country cannot be regarded as antisemitic. Antisemitism frequently charges Jews with conspiring to harm humanity, and it is often used to blame Jews for "why things go wrong." It is expressed in speech, writing, visual forms and action, and employs sinister stereotypes and negative character traits.*

*3.1 Contemporary examples of antisemitism in public life, the media, schools, the workplace, and in the religious sphere could, taking into account the overall context, include, but are not limited to:*

> *3.1.1 Calling for, aiding, or justifying the killing or harming of Jews in the name of a radical ideology or an extremist view of religion.*
>
> *3.1.2 Making mendacious, dehumanizing, demonizing, or stereotypical allegations about Jews as such or the power of Jews as collective — such as, especially but not exclusively, the myth about a world Jewish conspiracy or of Jews controlling the media, economy, government or other societal institutions.*
>
> *3.1.3 Accusing Jews as a people of being responsible for real or imagined wrongdoing committed by a single Jewish person or group, or even for acts committed by non-Jews.*
>
> *3.1.4 Denying the fact, scope, mechanisms (e.g. gas chambers) or intentionality of the genocide of the Jewish people at the hands of National Socialist Germany and its supporters and accomplices during World War II (the Holocaust).*

---

[21] https://www.holocaustremembrance.com/news-archive/working-definition-antisemitism

***3.1.5** Accusing the Jews as a people, or Israel as a state, of inventing or exaggerating the Holocaust.*

***3.1.6** Accusing Jewish citizens of being more loyal to Israel, or to the alleged priorities of Jews worldwide, than to the interests of their own nations.*

***3.1.7** Denying the Jewish people their right to self-determination, e.g., by claiming that the existence of a State of Israel is a racist endeavor.*

***3.1.8** Applying double standards by requiring of it a behavior not expected or demanded of any other democratic nation.*

***3.1.9** Using the symbols and images associated with classic antisemitism (e.g., claims of Jews killing Jesus or blood libel) to characterize Israel or Israelis.*

***3.1.10** Drawing comparisons of contemporary Israeli policy to that of the Nazis.*

***3.1.11** Holding Jews collectively responsible for actions of the state of Israel.*

***4.0 Antisemitic acts are criminal** when they are so defined by law (for example, denial of the Holocaust or distribution of antisemitic materials in some countries).*

***5.0 Criminal acts are antisemitic** when the targets of attacks, whether they are people or property – such as buildings, schools, places of worship and cemeteries – are selected because they are, or are perceived to be, Jewish or linked to Jews.]*

***6.0 Antisemitic discrimination** is the denial to Jews of opportunities or services available to others and is illegal in many countries.]*

# ANNEX II: INFERENCES OF THE IHRA WORKING DEFINITION

In order to apply the IHRA Working Definition of Antisemitism to the annotation of a dataset of tweets some parts require extrapolation. However, we strove to stay within the original meaning of the definition and refrained from adding any new concepts. The definition includes references, such as "classic antisemitism" and "stereotypical allegations about Jews." We made such references explicit by listing prominent stereotypes and images that are considered to be such stereotypes. We spell out explicitly what we believe is implicitly in the text. Any inferences should not be understood as changing the content of the definition.

The very first paragraph (1.0) notes that non-Jewish individuals can also become victims of antisemitism. We infer from section 5.0 that this is the case if they are perceived to be Jewish or linked to Jews. Additionally, we infer from section 3.1.2 that rhetoric can be antisemitic even if no specific Jewish individual or communal institution is targeted, but rather, the target is an abstract Jewish collective.

Section 3.1 lists 11 examples of contemporary forms of antisemitism. The IHRA has made it clear in additional statements that the examples are part of the definition.[22]

The example of Holocaust denial (3.14) includes denying the scope and the intentionality of the genocide of the Jewish people. Denying the scope of the Holocaust means denying that close to six million Jews were murdered for being Jews. The IHRA also adopted a "Working Definition of Holocaust Denial and Distortion." It is in accordance with its definition of antisemitism and further exemplifies that Holocaust denial "*may include publicly denying or calling into doubt the use of principal mechanisms of destruction (such as gas chambers, mass shooting, starvation and torture)*" and also "*blaming the Jews for either exaggerating or creating the Shoah for political or financial gain as if the Shoah itself was the result of a conspiracy plotted by the Jews.*"[23]

3.1.7 mentions denying the Jewish people their right to self-determination. Taking into account the next example, 3.1.8, "Applying double standards by requiring of it a behavior not expected or demanded of any other democratic nation" we include the denial of Israel's right to exist in its geographical region. However, the second part of 3.1.7, "claiming that the existence of a State of Israel is a racist endeavor" does not mean that all accusations of racism against Israel are antisemitic. It means that claiming that a State of Israel as per se racist (or an Apartheid state) is an example of denying the Jewish people their right to self-determination and is therefore antisemitic.

The Working Definition mentions "mendacious, dehumanizing, demonizing, or stereotypical allegations about Jews as such" and "classic stereotypes" without listing them explicitly. Below you find a composite of allegations and stereotypes that have become part of that repertoire. We compiled them by looking at descriptions that other scholars (Lipton 2014; Nirenberg 2013; Poliakov 2003b; 2003a; 2003c; 2003d; Rosenfeld 2013; 2015; Wistrich 2010; Perry and Schweitzer 2008; Livak 2010) have identified as prominent antisemitic stereotypes in the past 2000 years.

---

[22] A statement from July 19, 2018, on the IHRA website says: The Working Definition, including its examples, was reviewed and decided upon unanimously during the IHRA's Bucharest plenary in May 2016." https://www.holocaustremembrance.com/news-archive/working-definition-antisemitism A separate declaration was issued by UK delegates to the IHRA, August 7, 2018, due to a political debate about only partial adoption of the definition by the British Labor party. It included the following statement: "Any 'modified' version of the IHRA definition that does not include all of its 11 examples is no longer the IHRA definition." https://www.holocaustremembrance.com/news-archive/statement-experts-uk-delegation-ihra-working-definition-antisemitism

[23] The "Working Definition of Holocaust Denial and Distortion" was adopted by the IHRA's 31 member countries in October 2013, https://www.holocaustremembrance.com/working-definition-holocaust-denial-and-distortion.

Antisemitic allegations and stereotypes can be made by characterizing "the Jews" or by ascribing certain physical traits to them. Accusations of wrongdoing on the part of Jews also form part of the rich history of antisemitic stereotypes, as well as certain tropes. They can also be shown in the demonization of things and individuals that are thought of as being representative of Jews or Jewish beliefs. Certain beliefs, usually religious in nature, advocate for the punishment of Jews and also belong to the antisemitic tradition. Endorsing Nazism, Holocaust denial, or Israel-related forms of antisemitism are newer phenomena that are addressed explicitly and with examples in the working definition.

Supposed **"Jewish character"** is portrayed as stingy; greedy; immensely rich; being good with money; exploitative; corrupt; amoral; perverted; ruthless; cruel; heartless; anti-national/cosmopolitan; treacherous; disloyal; fraudulent; dishonest; untrustworthy; hypocrites; materialist; swank; work-shy; uncreative; intelligent; possess superhuman powers; arrogant; stubborn; culturally backwards; superstitious; ridiculous; dishonorable; hyper-sexual; ritually unclean; tribal; clannish; secretive; racist; men: effeminate and also lecherous; women: femme-fatal.

Supposed **"Jewish physical stereotypes"** are hooked noses; pointed beards; big ears; a weak or hunched frame; a dark complexion; hooves; horns; a tail and a goatee; unruly red or black *hair;* goggled eyes; blinded eyes; tired eyes; large lips; and an odor.

**Antisemitic imagery** can be found in depictions of Jews as the "wandering Jew"; demonic figures; lavishly rich capitalists; money/gold hoarding; hooked-nosed communists; heartless merchants; parasites and vile creatures such as beasts; octopi; snakes; rats; germs; and blood sucking entities.

Supposed "**Jewish crimes**" include the charge of deicide/ killing Jesus; being in league with the devil; seeking to destroy non-Jewish civilizations; working with alleged conspiratorial groups thriving for world power, such as Rothschilds, Freemasons, Illuminati, Jewish lobby, Zionist Lobby, Zionist Neocons, ZOG (Zionist Occupied Government); waging a (proxy) war against Islam/ Christianity; luring Christians/ Muslims away from Christianity/ Islam; profanation of Christian symbols; host desecration; practicing witchcraft; usury; profiteering; exploiting non-Jews; running transnational, allegedly "Jewish companies" in the interest of the Jews such as McDonalds, Starbucks, Coca Cola, Facebook; using blood from non-Jews for ritual purposes; killing or mutilating children for ritual purposes; adoring false gods and idols, such as the Golden Calf and Moloch; rejecting truth and being blind to the truth; perverting scripture; sticking to the letters but not the spirit of religious texts; falsifying scripture; having tried to murder the prophet Mohammed; well poisoning; causing epidemics, such as Black Death and AIDS; being responsible for the slave trade; poisoning non-Jews; aspiring to control the world secretly; secretly controlling world finance, country governments, media, Hollywood; orchestrating wars, revolutions, disasters (such as 9/11 and the subsequent wars in the Middle East); undermining culture and morals, especially concerning sexuality; degrading culture, music, science; degenerating race purity; undermining and betraying their countries of residence; inventing the Holocaust or exaggerating the Holocaust for material gain; being responsible for Christianity and the power of the church, for oligarchies, financial speculation, exploitation, capitalism, modernity, communism, bolshevism, liberalism, democracy, urbanization, Americanization, and globalization.

**Demonization** of things associated with Jews or of individuals seen as representative of Jews include the demonization of synagogues, Judaism, the Talmud, Kabbalah; the Judaization of enemies (using "Jew" as an insult); and the demonization of prominent Jews, such George Soros, Ariel Sharon, Benjamin Netanyahu, and Abraham Foxman as Jews.

**Nonvisual memes or recurrent phraseology** that are part of an antisemitic repertoire in different historical and cultural contexts include "Jews are the children/ spawn of Satan; synagogue of Satan; God has (eternally) cursed the Jews; Judaism (Jewish alleged choseness) is racist; Jews don't have a home country and cannot be a nation; Jews have no culture; Crypto Jews (converted Jews or their offspring remain Jewish and secretly act in the 'Jewish interest'); 'Jewish spirit' in science, music, culture is harmful to non-Jews; use of the terms 'Jewish terror' or 'Zydokumuna' for purges under communism; Jewish soldiers in WW1/ WW2 were traitors; all pro-Jewish or pro-Israeli organizations are funded/ operated by the Mossad; Jews are descendants of monkeys and pigs; Jews should never be taken as friends; Jews are the eternal enemies of Islam and Muslims. Muslims will kill the Jews at the end of time; reference to the battle of Khaybar; synagogues should be set on fire; Jews killed or sold out their own prophets/the son of God; Jews try to evade taxes/ Jews don't pay taxes; Hitler let some Jews live so that the world would know why he exterminated Jews; 'International Zionism' prevents a critical discussion about the Holocaust."

**Calls for punishment or justification of Jewish suffering** have also been part of an antisemitic discourse, mostly in religious contexts, such as "Jewish suffering is punishment by God; Humiliation and misery of Jews is proof of the truth of Christianity/Islam. Misery of Jews is proof of truth of Christianity.; Jews should be burnt as a form of punishment; Persecution of Jews under Hitler was punishment by God."

**Holocaust denial** is described in the IHRA Definition of Antisemitism. The more detailed IHRA Working Definition of Holocaust Denial and Distortion is used as additional guidance.

**Endorsing Nazism** today means endorsing the systematic killings of Jews by the Nazis (and their helpers). It is often done by the affirmative use of pro-Nazi memes and symbols.

**Manifestations of antisemitism related to Israel** are described in the Working Definition, including examples. Additional, frequently used antisemitic concepts include "Jews crucify or ritually kill Palestinians" and the use of term such as "Zionist Entity" to describe the State of Israel, which is a refusal to acknowledge the existence of Israel and thus a form of denying the Jewish people their right to self-determination. The following claims are "comparisons of contemporary Israeli policy to that of the Nazis" (example 3.1.10 in the working definition): equating Israeli politicians with Nazi leaders, such as Netanyahu = Hitler; claims that Israel engages in genocide/ "a Holocaust" against the Palestinian people; claims that the situation in the Gaza Strip is similar to the situation in the Warsaw Ghetto; using Nazi vocabulary to describe and denounce actions by the Israeli state, such as claims that Israel wages a war of extermination against the Palestinians.

Most **symbols that have an antisemitic connotation** are positive references to Nazism or to the Holocaust, such as the swastika or other Nazi or Nazi-predecessor flags, the Hitler salute, emblems of Nazi organizations such as the SS, Nazi slogans such as "Blut und Ehre" (blood and honor), or numbers representing "Heil Hitler" (88), "Adolf Hitler" (18). Positive references to the Holocaust used to manifest endorsement for the killing of Jews include symbols representing Zyklon B gas, including hissing noises, or references to ovens that were used to burn the people who were gassed (including pictures taken from these ovens).

The ADL Hate Symbols Database provides an extensive list of symbols used by hate groups, mostly white supremacists and Neo-Nazis.[24] This list helps us to contextualize tweets. We consider all symbols with positive references to Nazism antisemitic because they implicitly endorse the murder of Jews. Symbols by extremist Christian and Muslim groups are not as extensively documented (Ostovar 2017).[25] Some Jihadist groups, such as Hamas or Houthi,[26] are known for their antisemitism and lethally targeting victims as Jews. Endorsing the aforementioned groups is treated as a context suggesting antisemitism, but is annotated as antisemitic only when there are other references in the tweets to Jews, Judaism, or Israel that make an antisemitic reading of it likely.

---

[24] https://www.adl.org/education-and-resources/resource-knowledge-base/hate-symbols

[25] The Combating Terrorism Center at West Point has published an extensive list of Jihadist imagery with the Militant Imagery Project.

[26] The inscription of the Houthi flag reads (in Arabic): "God is the Greatest, Death to America, Death to Israel, Curse on the Jews, Victory to Islam."